# Sustainable regional economic development and land use: a case of Russia

**Wadim Strielkowski [1,2,*], Oxana Mukhoryanova [1], Oxana Kuznetsova [1] and Yury Syrov [1]**

[1] Institute of Economics and Management, North-Caucasus Federal University; Pushkin str. 1, 355017 Stavropol, Russian Federation; eac@ncfu.ru (W.S.); omukhorianova@ncfu.ru (O.M.); okuznetcova@ncfu.ru (O.K.); iusyrov@ncfu.ru (Y.S.)

[2] Department Agricultural and Resource Economics, University of California, Berkeley, 303 Giannini Hall, CA 94720, United States

* Correspondence: strielkowski@berkeley.edu

**Abstract:** This paper analyzes sustainable regional economic development and land use employing a case study of Russia. The economics of land management in Russia which is shaped by both historical legacies and contemporary policies represents an interesting conundrum. Following the dissolution of the Soviet Union, Russia embarked on a thorny and complex path towards the economic reforms and transformation characterized, among all, by the privatization and decentralization of land ownership. This transition was aimed at improving agricultural productivity and fostering sustainable regional economic development but also led to new challenges such as uneven distribution of land resources, unclear property rights, and underinvestment in rural infrastructure. However, managing all of that effectively poses significant challenges and opportunities. With the help of the comprehensive bibliographic network analysis, this study sheds some light on the current state of sustainable regional economic development and land use management in Russia. Its results and outcomes might be helpful for the researchers and stakeholders alike in devising effective strategies aimed at maximizing resources for sustainable land use, particularly within their respective regional economies.

## 1. Introduction

In recent years, there has been a renewed focus on improving land use management practices to address the issues of sustainable economic development and regional development all over the world, making this a universal trend [1, 2]. The Russian Federation has also initiated several reforms aimed at enhancing land tenure security, promoting sustainable agricultural economic practices, and encouraging investment in rural areas [3, 4]. Despite these efforts, progress remains uneven across different regions due to varying levels of economic development, environmental conditions, and administrative capacities. One critical aspect currently characterizing Russian land management is its approach toward environmental sustainability [5-7]. With climate change posing an ever-increasing threat to Russia's ecosystems, from its vast taiga forests to its Arctic territories, there is an urgent need for integrating ecological considerations into land use planning. Efforts are being made towards adopting more sustainable agricultural techniques that minimize soil degradation and water use while preserving biodiversity [8]. Another significant factor influencing land management in Russia is technological innovation [9-11]. Digital technologies such as satellite imaging and geographic information systems are increasingly being utilized for efficient monitoring and planning purposes. These technologies offer potential for improved decision-making processes concerning land allocation, risk assessment related to natural disasters, and tracking changes in land use patterns over time [12, 13]. However, realizing the full potential of these innovations requires overcoming existing barriers such as limited access to technology in remote areas, lack of skilled personnel familiar with advanced technological tools, and resistance from many traditional sectors reluctant to adopt new practices [14, 15].

Thence, understanding the current state of land use and land management in Russia reveals a complex picture marked by both achievements and ongoing challenges. While there are clear paths forward through policy reformulation aimed at sustainability goals coupled with technological integration for better governance structures as far as achieving a holistic transformation necessitates concerted efforts across governmental levels [16, 17}. This collective approach would ensure that strategies devised not only maximize resources but also pave the way towards a sustainable future for Russia's regional economies amidst changing global environmental dynamics [18].

**2. Strategies for sustainable land use: an overview of strengths and weaknesses**

Maximizing resources through strategies for sustainable land use is an imperative approach to ensure the economic success and overall resilience of Russia's regional economies. The vast territory of the Russian Federation, with its diverse climates, topographies, and ecosystems, presents unique opportunities and challenges for sustainable land management [19, 20]. Table 1 that follows captures how the strategies for sustainable land use in Russia leverage the nation's vast natural resources and geographical diversity while facing challenges related to implementation scale, coordination, and integration of new technologies with traditional practices.

**Table 1.** Overview of strengths and weaknesses of maximizing resources through strategies for sustainable land use in Russia

| Aspect | Strengths | Weaknesses |
| --- | --- | --- |
| *Economic and ecological balance* | Balances economic development with ecological conservation, integrating both goals in regional strategies. | Requires complex coordination to maintain balance without sacrificing one for the other. |
| *Agricultural innovation* | Precision agriculture optimizes resource use and enhances productivity while preserving soil health. | High initial investment and technical expertise needed to implement and maintain advanced agricultural technologies. |
| *Agroforestry* | Creates sustainable land-use systems that combine agriculture and forestry, increasing diversity and productivity. | May be limited by climatic and topographical constraints; needs tailored approaches for different regions. |
| *Forest management* | Supports biodiversity, carbon sequestration, and sustainable industries through practices like selective logging and reforestation. | Ongoing monitoring and enforcement needed to prevent illegal activities and ensure practices are sustainable. |
| *Urban planning* | Mitigates urban sprawl and enhances living conditions through efficient space use and green infrastructure. | Urban planning reforms can be slow to implement and require significant alterations in existing infrastructure and planning policies. |
| *Conservation policies* | Protects vulnerable ecosystems and ensures responsible resource extraction with environmental impact assessments. | Requires rigorous and sometimes costly assessment processes; potential resistance from industries like mining or oil extraction heavily invested in certain regions. |
| *Cooperation and compliance* | Involves cooperation between multiple governance levels and sectors, encouraging broad adoption of sustainable practices with clear guidelines and incentives. | Coordination challenges across vast territories with diverse local needs; compliance may vary due to regional governance quality or economic pressures. |
| *Technological and traditional integration* | Combines modern technologies with traditional knowledge, fostering innovation while preserving indigenous practices and ecological knowledge. | Integrating vastly different approaches can be challenging, and there may be resistance to new technologies or methodologies in certain communities |

Source: Own results

Hence, sustainable land use and land management both not only encompass the preservation of natural resources but also involve optimizing these resources in a manner that does not compromise the ability of future generations to meet their needs [21, 22]. The cornerstone of sustainable land use strategy in Russia's regional economy lies in balancing economic development with ecological conservation. This involves a multifaceted approach that integrates agricultural innovation, forest management, urban planning, and conservation policies [23, 24]. By adopting practices such as precision agriculture and agroforestry, regions can enhance productivity while minimizing environmental impacts. Precision agriculture utilizes advanced technologies to ensure that crops receive precisely what they need for optimal growth thereby reducing waste and preserving soil health. Agroforestry, on the other hand, combines agriculture and forestry to create more diverse, productive, sustainable land-use systems [25].

Moreover, forest management plays a crucial role in Russia's sustainability efforts given its status as home to the world's largest forest reserves. Sustainable forestry practices such as selective logging and reforestation are essential for maintaining biodiversity, sequestering carbon dioxide, and supporting industries dependent on timber and other forest products without depleting these valuable resources [26, 27].

In addition, urban planning also contributes significantly to sustainable land use by reducing sprawl and encouraging more efficient uses of space within Russia's rapidly growing smart cities [28-30]. Implementing green infrastructure like parks or green roofs can mitigate urban heat island effects while enhancing quality of life for residents [31, 32].

Furthermore, strategic zoning laws can protect valuable agricultural lands from being overtaken by urban development. Another critical aspect is fostering conservation policies that protect vulnerable ecosystems while promoting responsible resource extraction where appropriate. This requires rigorous environmental impact assessments before initiating projects like mining or oil extraction which are prominent in certain Russian regions [33, 34]. To effectively implement these strategies across Russia's vast territories requires robust cooperation between federal authorities and regional governments along with active participation from local communities and businesses. Establishing clear guidelines for sustainable practices along with incentives for compliance can encourage broader adoption across sectors [35].

Therefore, maximizing resources through strategies for sustainable land use is pivotal for securing the future prosperity of Russia's regional economies. It necessitates a holistic approach that embraces technological innovation alongside traditional knowledge while prioritizing ecological integrity alongside economic objectives—an equilibrium essential for sustaining both people's livelihoods and planet's well-being over time.

**3. Economic development in promoting sustainable land use**

The role of economic development in promoting sustainable land use within Russia's expansive and diverse regional economy cannot be overstated [36, 37]. As the world grapples with the pressing need for sustainability, Russia, endowed with vast natural resources and significant agricultural potential, finds itself at a pivotal juncture. The intersection of economic development and sustainable land use in this context is critical, not only for ensuring long-term ecological balance but also for fostering regional economic prosperity. Economic development in Russia's regions has traditionally been driven by resource extraction and heavy industry. While these sectors have contributed significantly to the country's GDP and employment rates, they have also led to considerable environmental degradation and inefficient land use practices [38-40]. Therefore, the challenge lies in reorienting economic strategies towards models that prioritize sustainability without compromising growth.

Therefore, promoting sustainable land use through economic development involves a multifaceted approach that intertwines technological innovation, policy reform, and community engagement. One of the key strategies is the adoption of advanced agricultural techniques that increase productivity while minimizing environmental impact [41-43]. Precision farming, crop rotation, and soil conservation methods can enhance yield per hectare, reducing the need for expanding agricultural lands into natural habitats. Moreover, diversifying regional economies beyond resource extraction is vital. Developing sectors such as renewable energy, eco-tourism, and organic agriculture not only offers alternative sources of income but also incentivizes conservation efforts [44-46]. For instance, regions abundant in natural beauty or biodiversity can leverage these assets through eco-tourism ventures that generate revenue while fostering an appreciation for environmental stewardship among both locals and visitors [47]. Policy plays a crucial role in aligning economic incentives with sustainable outcomes. Implementing regulations that limit harmful industrial practices or promote green technologies can steer businesses towards more responsible operations. Financial mechanisms such as subsidies for sustainable practices or taxes on carbon emissions can further encourage compliance with environmental standards [48, 49].

However, achieving sustainable land use through economic development requires more than just top-down interventions. It demands active participation from local communities. Empowering residents to manage their resources sustainably ensures that development projects are tailored to specific regional needs and conditions. This participatory approach fosters a sense of ownership among community members over their environment and its preservation [50].

Hence, integrating sustainability into Russia's regional economic development agenda is imperative for securing both ecological integrity and socio-economic resilience. By embracing innovative agricultural practices, diversifying economies away from unsustainable industries, enacting supportive policies, and engaging local communities, the country can pave the way towards a future where prosperous economies coexist harmoniously with healthy ecosystems. Such strategic alignment between economic development objectives and sustainable land use principles holds the promise of transforming Russia's regions into models of balanced prosperity for generations to come.

Table 2 that follows summarizes the main points of the role of economic development in promoting sustainable land use in Russia.

**Table 2.** Role of economic development in promoting sustainable land use in Russia

| Factors | Impact |
| --- | --- |
| **Context and importance** | Economic development is crucial for sustainable land use in Russia, a country with vast natural resources and significant agricultural potential. This intersection is vital for ecological balance and regional prosperity. |
| **Current economic drivers** | Traditional economic activities like resource extraction and heavy industry have boosted GDP and employment but have also led to environmental degradation and inefficient land use. |
| **Challenges** | The main challenge is shifting from traditional economic models to sustainable practices without hindering growth. |
| **Strategic approaches** | Economic development should include technological innovation, policy reform, and community engagement to promote sustainable land use. |
| **Agricultural innovations** | Adoption of advanced techniques like precision farming and soil conservation to increase productivity and reduce the expansion into natural habitats. |
| **Economic diversification** | Diversifying from resource-heavy industries to sectors like renewable energy, eco-tourism, and organic agriculture can provide sustainable income sources and promote conservation. |
| **Policy implementation** | Policies should align economic incentives with sustainable outcomes, utilizing regulations and financial mechanisms like subsidies for green practices and taxes on carbon emissions. |
| **Community engagement** | Local community involvement is essential for tailoring development to regional needs and fostering ownership and commitment to environmental stewardship. |

Source: Own results

## 4. Materials and methods

In the empirical part of our study, we employ bibliometric analysis that is based on a dataset of 377 documents (represented by research papers, conference proceedings, and book chapters) indexed in the Web of Science (WoS) database between 1991 and 2024.

We have picked the Web of Science (WoS) database as being one of the most prestigious and complete abstract and citation databases that exist and offer a vast body of relevant literature on the topic of our research.

From Figure 1 that follows, one can see the growing trend in the number of publications on land use, regional economic development, as well as land management related to the Russian Federation in the recent years even though some varying peaks and slumps are visible in the dissemination during certain periods (e.g. during and after the COVID-19 pandemic) (see Figure 1). The year 2024 has been left out due to the fact that the number of publications is still being accumulated and is not precise at the moment.

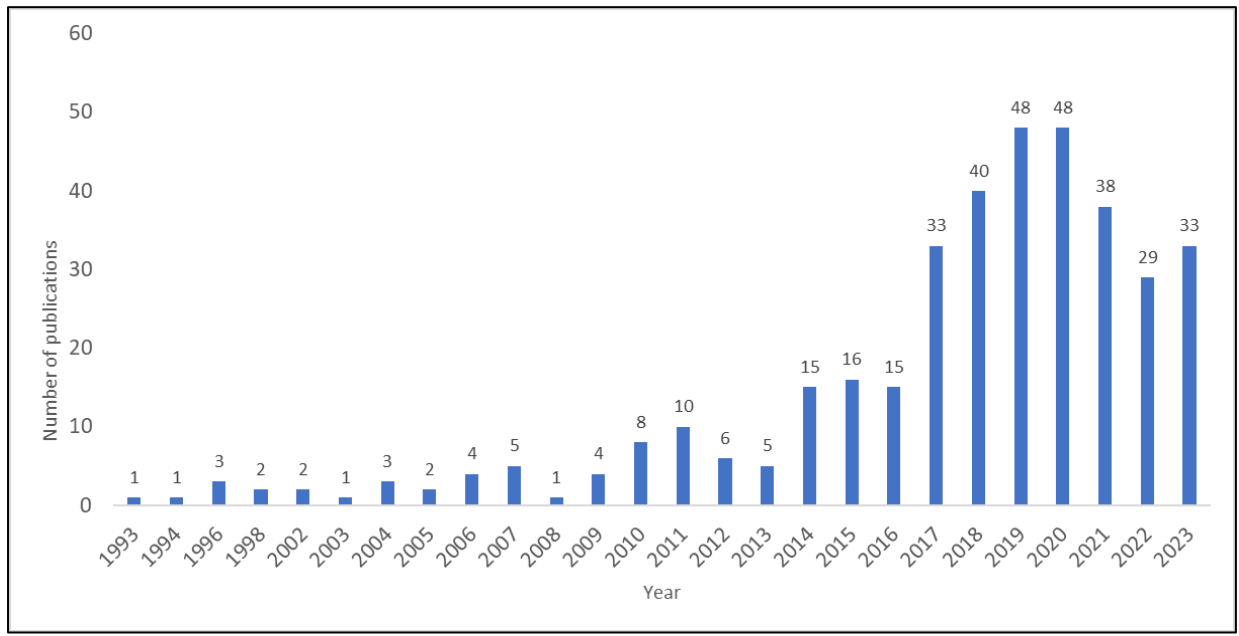

**Fig. 1.** Annual number of publications land use, regional economic development, as well as land management in Russia (1991-2023).
Source: Own results based on WoS.

Using the set of data described above, we conducted a network cluster analysis of the textual and bibliometric inputs using VOSViewer software. Figure 2 that follows provides the description of our algorithm for the data selection, retrieval, processing, as well as the network analysis used in our study. The diagram delineates the methodological approach for data selection and network analysis in a stepwise fashion. Initially, the research phenomenon is defined, which informs the choice of the Web of Science (WoS) as the database for sourcing research publications. The search criteria are then established through the searches in the WoS database for "land use" AND "Russian Federation" covering a timeframe from 1991 to 2024.

The search within the chosen database is predetermined by some specific criteria, including the scope (title, abstract, and keywords), the timeframe (1991 to 2024), language, the quality of the studies, and the type of document, which includes articles, conference proceedings, as well as book chapters. The analysis considers the distribution of documents by year, affiliations, journals, and institutions. The extracted data is then inputted into the VOSviewer software, a tool designed for constructing and visualizing bibliometric networks.

Our analysis of the results generated by the VOSviewer software includes network and overlay visualization, the identification of the number of clusters, types of clusters, and the total link strength. All of these involve interpreting and evaluating the implications stemming from the network analysis. The whole process concludes with the synthesis of the findings that allow us to draw conclusions and implications from the study backed up by the results of the bibliometric analysis. This robust comprehensive approach ensures a systematic and rigorous analysis of the data, facilitating a clear and precise understanding of the studied research phenomena and their implications and outcomes (see Figure 2 below).

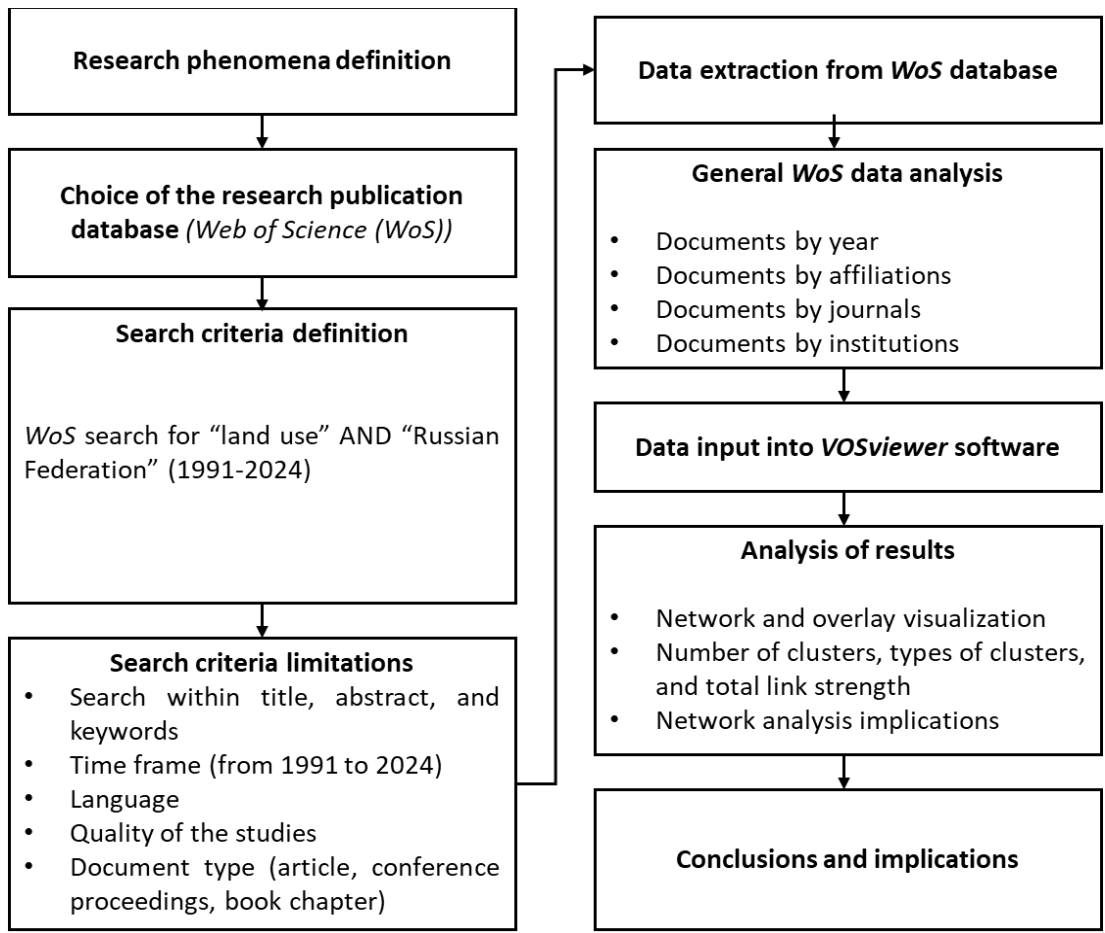

**Fig. 2.** Diagram showing the data selection and network analysis algorithm.
Source: Own results

**5. Main results**

Figure 3 that follows depicts the co-occurrence map based on the text data 377 papers containing the keywords "land use" and "Russian Federation" retrieved from WoS database. Our analysis has identified just two main clusters (marked in green and red colors).

The keywords in the green cluster seem to be more focused on environmental and practical aspects of land use, such as "soil," "forest," "agriculture," "area," and "territory." These could represent papers that are concerned with the ecological, agricultural, and spatial dimensions of land use. The red cluster includes terms like "law," "legislation," "right," "ownership," and "land legislation." This indicates a legal and regulatory focus within the dataset, where papers are likely discussing the legal frameworks, property rights, and legislative aspects of land use.

It is obvious that there exist numerous links between the two clusters, showing that there is a significant degree of interrelation between the practical/environmental and legal/regulatory aspects of land use in the Russian Federation. This suggests that the literature recognizes the importance of legal frameworks in shaping and managing the ecological and practical use of land.

In Figure 3, one can notice that certain keywords like "Russia," "land use," and "assessment" are more centrally located and have many connections to other terms, which implies these are key topics within the research literature and may serve as bridging concepts between different thematic areas.

In addition, the specific mention of terms like "land plot," "case," "article," and "concept" suggests detailed discussions in the literature, potentially regarding specific legal cases, articles of law, or conceptual debates about land use.

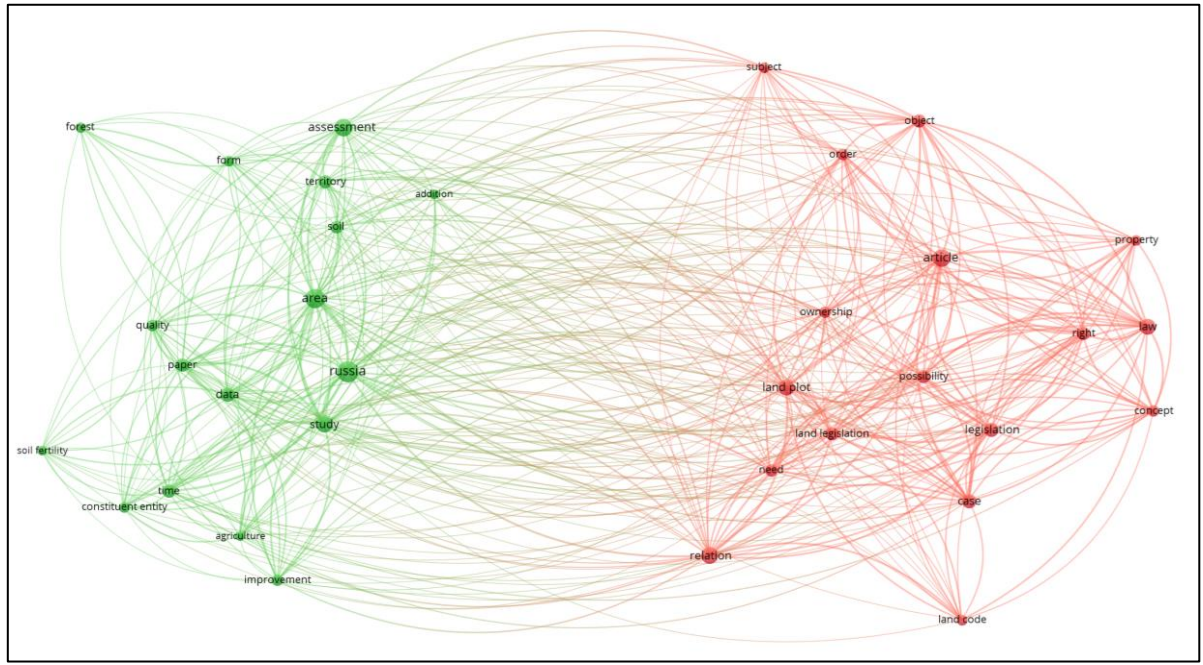

**Figure 3.** Co-occurrence map based on the text data of the 377 papers containing the keywords "land use" and "Russian Federation" retrieved from WoS database.
Own results based on VOSViewer v.1.6.19 software

Furthermore, keywords at the periphery, such as "soil fertility," "constituent entity," and "improvement" indicate niche areas of research that are not as extensively connected to the core topics but still relevant within the dataset.

Based on this analysis, one can infer that the research landscape for "land use" in the "Russian Federation," is multidimensional, addressing both the practical management of land and the overarching legal and regulatory frameworks. The presence of strong interconnections between clusters emphasizes the interdisciplinary nature of land use research, where environmental management and policy/legal frameworks are intertwined.

Figure 4 that follows below depicts the bibliographic map based on the bibliometric data (co-authorship, keyword co-occurrence, citation, bibliographic coupling, or co-citation map) of the 377 papers containing the keywords "lad use" and "Russian Federation" in WoS database (see Figure 4).

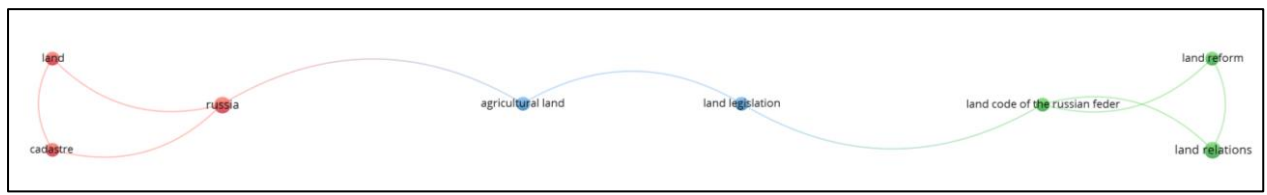

**Figure 4.** Bibliographic map based on the bibliometric data (co-authorship, keyword co-occurrence, citation, bibliographic coupling, or co-citation map) 377 papers containing the keywords "land use" and "Russian Federation" in WoS database.
Own results based on VOSViewer v.1.6.19 software

In Figure 4, there are just two clusters identified by color (red on the left and green on the right). This could indicate a progression or evolution of the topics from more general concepts (in red) to more specific themes or applications (in green). The red cluster begins with terms like "land" and "Russia," and includes "cadastre," which refers to a comprehensive land recording system. The presence of these terms together suggests that the initial focus in the literature may be on the foundational aspects of land use such as land registration and the documentation processes in Russia. It is noteworthy that the terms

"agricultural land" sits at the transition between the red and green clusters. This indicates that discussions in literature might evolve from general land use and cadastre to more specific applications in the context of agriculture.

The terms in the green cluster are more specific and are related to legal and policy aspects such as "land legislation," "land code of the Russian Federation," "land reform," and "land relations." This progression signifies a focus on the legal framework governing land use, reforms to this framework, and the relationships and regulations around land ownership and use.

The curves connecting the clusters show relationships between the terms across the clusters. The more pronounced curve between "land" and "land relations" suggests that there is a significant connection between general land use issues and the specific legal relations governing land. The increasing specificity of keywords from left to right might reflect a progression in research focus, from general descriptions of land and cadastre systems to detailed discussions of legal frameworks and reforms.

Based on Figure 4, it is possible to conclude that the dataset of selected papers picked from WoS database primarily covers two thematic areas: foundational concepts of land and cadastral records, and more specialized discussions on the legal aspects of land use in the Russian Federation. The linear and progressive nature of the map suggests that there might be a developmental pathway in the literature from general land registration systems to detailed legal reforms and relations. The visual simplicity of the network indicates a clear thematic direction in the research topics, rather than a complex interweaving of themes.

## 6. Conclusions

All in all, our comprehensive results stemming from the literature review and the bibliometric network analysis unveiled a comprehensive landscape of research topics, demonstrating the multidimensional nature of land management in Russia. It becomes evident that sustainable land use is contingent upon the synergistic interplay between environmental management and the legal-regulatory framework.

Our findings highlight that despite the challenges, there is an existing framework within Russia for balancing economic development with ecological conservation. This is exemplified by the dual focus on enhancing agricultural productivity through innovations such as precision agriculture and agroforestry, and on protecting ecological diversity via sustainable forest management and urban planning.

Crucially, the results underscore the importance of policy reforms that prioritize sustainability while fostering economic growth. The research has shown that shifts towards sustainable practices are achievable through strategic policy reforms, technological integration, and engaging local communities in the management of land resources.

The study's outcomes suggest several policy implications that could foster sustainable land use in Russia: First of all, there appears to be a need for policies promoting wider access to digital technologies across all regions, ensuring that advanced tools for land use planning and risk assessment are universally available. Second, encouraging the development of skills necessary to utilize advanced technologies should be a policy priority. This includes training and educational programs focusing on precision agriculture, sustainable forestry management, and urban planning. Third, policies should incentivize diversification of regional economies, moving away from reliance on resource extraction to sustainable sectors like renewable energy, eco-tourism, and organic agriculture. Fourth, intergovernmental coordination is paramount for successful policy implementation. Policies must encourage cooperation across federal, regional, and local levels, alongside engagement with private sectors and local communities. Fifth, strengthening the legal framework surrounding land rights, property ownership, and resource extraction is essential. This includes clear land legislation that supports sustainable land use and efficient dispute resolution mechanisms. Finally, conservation policies require reinforcement through regulations and financial instruments that encourage compliance with environmental standards, alongside active conservation efforts.

Our study has shed considerable light on the current state of land management in Russia, but it has also opened several pathways for further research. There clearly exists a need for long-term studies examining the effects of policy changes on land use patterns, assessing the effectiveness of current sustainability initiatives over time. In addition, comparative research across different regions within Russia could elucidate how various economic, social, and environmental factors influence the success of sustainable land use practices. Moreover, evaluating the impacts of technological interventions and policy reforms on both economic development and environmental sustainability would provide deeper insights into their effectiveness and areas for improvement. Research into successful models of community engagement in land management could help in formulating strategies that foster more participatory approaches. Further exploration into how indigenous practices and local ecological knowledge can be integrated into modern sustainable land management practices would be beneficial. Additionally. investigating the barriers to the adoption of technological innovations in land use planning and management can inform policies that address these impediments. Finally, comparing Russia's experience with other countries facing similar challenges could also yield valuable lessons and innovative strategies that could be adapted and implemented.

**Acknowledgements:** The research was supported by the Russian Science Foundation, project No. 23-28-10154, https://rscf.ru/project/23-28-10154/.

**Conflicts of Interest:** The authors declare no conflict of interest.